\documentclass{article}
\usepackage{epsfig}
\usepackage{amsmath}
\newcommand{\e}{\text{e}}
\numberwithin{equation}{section}
\begin{document}
\title{NON-ABELIAN CONSTRAINTS\\ IN MULTIPARTICLE PRODUCTION
\thanks{Presented at the Cracow Epiphany Conference on Quarks and Gluons in Extreme Conditions, Cracow, Poland, January 3-6, 2002}}%
\author{Ludwik Turko\thanks{turko@ift.uni.wroc.pl}\\
\small Institute of Theoretical Physics, University of Wroclaw,\\
\small Pl. Maksa Borna 9, 50-204 Wroc\l aw, Poland }
\date{}
\maketitle
\begin{abstract}
Internal microscopic symmetry of a many body system leads
to global constraints. We obtain explicit forms of the global
macroscopic condition assuring that at the microscopic level  the
evolution respects the overall symmetry.
\end{abstract}
\section{Introduction}

Let us consider a multiparticle interacting system with the internal symmetry
taken into account. The internal symmetry leads to conservation laws which put
constraints on the evolution of system. The problem arises if are there only
constraints due to the symmetry conservation? In the microscopic formulation
with symmetry invariant dynamical equations the answer is given by an analysis
of corresponding solutions --- assuming that solutions are known.

We are looking here for global conditions to provide consistency with the
overall symmetry of the system. These conditions should not depend on the
exact analytic form of the solutions. As an example one can take Kepler's laws
in the classical mechanics which are related to the orbital momentum
conservation and can be proved without knowledge of the analytic solutions of
Newton equations. Another simple example is a case of $n$ species of charged
particles with individual charges $q_a, q_b,\dots, q_n.$ Numbers of particles
a given by $N^{(a)}, N^{(b)},\dots, N^{(n)}.$

Particle numbers are time-dependent but the global charges must be
conserved (exact $U(1)$ symmetry). So there is a condition
\begin{equation}\label{abcharge}
q_a\frac{d N^{(a)}}{dt} + q_b\frac{d N^{(b)}}{dt}+ \cdots + q_n\frac{d
N^{(n)}}{dt} = 0\,,
\end{equation}
valid for any charge conserving interaction.

Our aim\cite{TurRaf} is to find a corresponding condition for non-abelian
symmetries. In the non-abelian case there is a subsidiary condition besides
conditions of the type Eq.\, (\ref{abcharge}) due to the charge conservation.
This is the demand to preserve the internal symmetry group representations
during the evolution of the system.

\section{Generalized Projection Method}

Let us consider a system big enough to use methods of statistical physics.
When the system reaches the statistical equilibrium then one can extract
contributions from particular irreducible representation of the symmetry group
\cite{TurRed}. Group projection techniques allowed for a consistent treatment
of equilibrium systems and gave tools to obtain canonical partition functions
corresponding to the system transforming under given representation of the
symmetry group. This technique can also be used for a more general non-static
problem.

Let us consider a system consisting of particles belonging to
multiplets $\alpha_j$ of the symmetry group. Particles from the
given multiplet $\alpha_j$ are characterized by quantum numbers
$\nu_j$ --- related to the symmetry group, and quantum numbers
$\zeta_j$ characterizing different multiplets of the same irreducible
representation $\alpha_j$.

The number of particles of the specie $\{\alpha,\nu_\alpha;\zeta\}$ is denoted
here by $N^{(\alpha)}_{\nu_\alpha;(\zeta)}.$ These occupation numbers are time
dependent until the system reaches the chemical equilibrium. However, the
representation of the symmetry group for the system remains constant in the
course of a time evolution. A multiplicity $N^{(\alpha_j)}$ of the
representation $\alpha_j$ in this product is equal to a number of particles
which transform under this representation
\begin{equation}
N^{(\alpha_j)} = \sum_j\left(\sum_{\zeta_j}\,
N^{(\alpha_j)}_{\nu_{\alpha_j};(\zeta_j)}\right)= \sum_j\,
N^{(\alpha_j)}_{\nu_{\alpha_j}}\,.
\end{equation}
We introduce a state vector $\left\vert
N^{(\alpha_1)}_{\nu_{\alpha_1}},\dots,
N^{(\alpha_n)}_{\nu_{\alpha_n}}\right\rangle $ in particle number
representation. The probability that
$N^{(\alpha_1)}_{\nu_{\alpha_1}},\dots, N^{(\alpha_n)}_{
\nu_{\alpha_n}}$ particles transforming under the symmetry group
representations $\alpha_1,\dots,\alpha_n$ combine into a state
transforming under representation $\Lambda$ of the symmetry group
is given by
\begin{eqnarray}\label{proj1states}
\overline{P^{\Lambda,\lambda_{\Lambda}}_{\{N^{(\alpha_1)}_{\nu_{\alpha_1}},
\dots,\,N^{(\alpha_n)}_{\nu_{\alpha_n}}\}}}\quad =
\quad\left\langle N^{(\alpha_1)}_{\nu_{\alpha_1}},
\cdots,N^{(\alpha_n)}_{\nu_{\alpha_n}}\right\vert
{\mathcal P}^{\Lambda}\left\vert N^{(\alpha_1)}_{\nu_{\alpha_1}},
\dots, N^{(\alpha_n)}_{\nu_{\alpha_n}}\right\rangle
\end{eqnarray}
The projection operator ${\mathcal P}^{\Lambda}$ has the form (see
\textit{e.g.} \cite{Wigner})
\begin{equation}\label{proj1}
{\mathcal
P}^{\Lambda}=d(\Lambda)\int\limits_G\,d\mu(g)\bar\chi^{(\Lambda)}(g)U(g)\,.
\end{equation}
Here $\chi^{(\Lambda)}$ is the character of the representation $\Lambda$,
$d(\Lambda)$ is the dimension of the representation, $d\mu(g)$ is the invariant
Haar measure on the group, and $U(g)$ is an operator transforming a state under
consideration. In particle number representation the operator $U(g)$ is defined
as
\begin{eqnarray}\label{transf}
\lefteqn{U(g)\left\vert N^{(\alpha_1)}_{\nu_{\alpha_1}},\dots,
N^{(\alpha_n)}_{\nu_{\alpha_n}}\right\rangle}\\ & &
=\sum\limits_{\nu_1^{(1)},\dots,\nu_n^{(N_{\nu_n})}}\,
D^{(\alpha_1)}_{\nu_1^{(1)}\nu_1}\!\!\cdots
D^{(\alpha_1)}_{\nu_1^{(N_{\nu_1})}\nu_1}\!\!\cdots
D^{(\alpha_n)}_{\nu_n^{(1)}\nu_n}\cdots
D^{(\alpha_n)}_{\nu_n^{(N_{\nu_n})}\nu_n} \left\vert
N^{(\alpha_1)}_{\nu_{\alpha_1}},\dots,
N^{(\alpha_n)}_{\nu_{\alpha_n}}\right\rangle\,.\nonumber
\end{eqnarray}
$D^{(\alpha_n)}_{\nu,\nu}$ is a matrix elements of the group
element $g$ corresponding to the representation $\alpha$.

One gets finally
\begin{eqnarray} \label{weights}
\overline{P^{\Lambda,\lambda_{\Lambda}}_{\{N^{(\alpha_1)}_{\nu_{\alpha_1}},
\dots,\,N^{(\alpha_n)}_{\nu_{\alpha_n}}\}}}\\ & & = {\mathcal A}^{\{{\mathcal
N}\}} d(\Lambda)\int\limits_G\,d\mu(g)\bar\chi^{(\Lambda)}(g)
[D^{(\alpha_1)}_{\nu_1\nu_1}]^{N^{(\alpha_1)}_{\nu_{\alpha_1}}} \cdots
[D^{(\alpha_n)}_{\nu_n\nu_n}]^{N^{(\alpha_n)}_{\nu_{\alpha_n}}}\,.\nonumber
\end{eqnarray}
$D^{(\alpha_n)}_{\nu,\nu}$ is a matrix elements of the group element $g$
corresponding to the representation $\alpha$ and ${\mathcal A}^{\{{\mathcal
N}\}}$ is an overall permutation normalization factor
\begin{equation}\label{faktor}
{\mathcal A}^{\{{\mathcal N}\}}=
\prod\limits_j\prod_{\zeta_j}{\mathcal A}^{\alpha_j}_{(\zeta_j)}\,,
\end{equation}
where ${\mathcal A}^{\alpha_j}_{(\zeta_j)}$ are partial factors for
particles of the kind $\{\alpha,\zeta\}:$
\begin{eqnarray}\label{faktorp}
{\mathcal A}^\alpha_{(\zeta)}= \frac{{\mathcal
N}^{(\alpha)}_{(\zeta)}!}{d(\alpha)^{{\mathcal
N}^{(\alpha)}_{(\zeta)}}\prod\limits_{\nu_\alpha} {\mathcal
N}^{(\alpha)}_{\nu_\alpha;(\zeta)}!}\,.
\end{eqnarray}
The permutation factor gives a proper normalization of state vectors reflecting
indistinguishability of particles
\begin{equation}\label{norma}
\left\langle N^{(\alpha_1)}_{\nu_{\alpha_1}},
\cdots,N^{(\alpha_n)}_{\nu_{\alpha_n}}\right\vert \left.
N^{(\alpha_1)}_{\nu_{\alpha_1}},\dots,
N^{(\alpha_n)}_{\nu_{\alpha_n}}\right\rangle={\cal A}^{\{{\cal N}\}}\,.
\end{equation}
Because of the symmetry conservations all weights in Eq.\,(\ref{weights})
should be constant
\begin{equation} \label{cond}
\frac{d}{dt}\overline{P^{\Lambda,\lambda_{\Lambda}}_{\{N^{(\alpha_1)}_{\nu_{\alpha_1}},
\dots,\,N^{(\alpha_n)}_{\nu_{\alpha_n}}\}}}\quad =\quad 0 .
\end{equation}
Introducing  here the result of Eq.\,(\ref{weights}) one obtains

\begin{eqnarray}\label{deriv}
 0&=&\frac{d\,\log{\mathcal A}^{\{{\mathcal N}\}}}{dt}\int\limits_G
d\mu(g)\bar\chi^{(\Lambda)}(g)
[D^{(\alpha_1)}_{\nu_1\nu_1}]^{N^{(\alpha_1)}_{\nu_{\alpha_1}}}\cdots
[D^{(\alpha_n)}_{\nu_n\nu_n}]^{N^{(\alpha_n)}_{\nu_{\alpha_n}}}\\
&+&\sum_{j=1}^n \sum_{\nu_{\alpha_j}}\,
\frac{d\,N^{(\alpha_j)}_{\nu_{\alpha_j}}}{dt}\int\limits_G
d\mu(g)\bar\chi^{(\Lambda)}(g)
[D^{(\alpha_1)}_{\nu_1\nu_1}]^{N^{(\alpha_1)}_{\nu_{\alpha_1}}} \cdots
[D^{(\alpha_n)}_{\nu_n\nu_n}]^{N^{(\alpha_n)}_{\nu_{\alpha_n}}}
\log[D^{(\alpha_j)}_{\nu_j\nu_j}]\nonumber\,.
\end{eqnarray}

The integrals which appear in Eq.\,(\ref{deriv}) can be expressed explicitly in
an analytic form for any compact symmetry group.

To write an expression for the time derivative of the normalization factor
${\mathcal A}^{\{{\mathcal N}\}}$ we perform analytic continuation from integer
to continuous values of variables $N^{(\alpha_n)}_{\nu_{\alpha_n}}.$ All
factorials in Eq.\, (\ref{faktor}) are replaced by the $\Gamma$--function  of
corresponding arguments.  We encounter here also the digamma function $\psi$
\cite{Abram}
\begin{equation}\label{digamma}
\psi(x)=\frac{d\, \log\Gamma(x)}{d\,x} .
\end{equation}
This allows to write
\begin{multline} \label{evfactor}
\frac{d\,\log{{\mathcal A}^{\{{\mathcal N}\}}}}{dt}\\ =
  \sum_j\sum_{\zeta_j}
 \left[\frac{d\,{\mathcal N}^{(\alpha_j)}_{(\zeta_j)}}{dt}
 \psi({\mathcal N}^{(\alpha_j)}_{(\zeta_j)}+1)
 -\sum_{\nu_{\alpha_j}}
\frac{d\,{\mathcal N}^{(\alpha_j)}_{\nu_{\alpha_j};(\zeta_j)}}{dt}
\psi({\mathcal N}^{(\alpha_j)}_{\nu_\alpha;(\zeta_j)}+1)\right]\,.
\end{multline}
Eq.\,(\ref{deriv}) can be written in a form
\begin{equation}\label{derisimpl}
\begin{split}
\sum_{j=1}^n \sum_{\nu_{\alpha_j}}\,&
\frac{d\,N^{(\alpha_j)}_{\nu_{\alpha_j}}}{dt}\frac{d\,\log \widetilde
\mathcal{P}^{\Lambda,\lambda_{\Lambda}}_{\{N^{(\alpha_1)}_{\nu_{\alpha_1}},
\dots,\,N^{(\alpha_n)}_{\nu_{\alpha_n}}\}}}
{d\,N^{(\alpha_j)}_{\nu_{\alpha_j}}}\\
= \sum_j\sum_{\zeta_j}&
\left(-\frac{d\,N^{(\alpha_j)}_{(\zeta_j)}}{dt}
\psi(N^{(\alpha_j)}_{(\zeta_j)}+1)+\sum_{\nu_{\alpha_j}}
\frac{d\,N^{(\alpha_j)}_{\nu_{\alpha_j};(\zeta_j)}}{dt}
\psi(N^{(\alpha_j)}_{\nu_{\alpha_j};(\zeta_j)}+1)\right)\,,
\end{split}
\end{equation}
where
\begin{equation}\label{otherP}
\widetilde {\mathcal
P}^{\Lambda,\lambda_{\Lambda}}_{\{N^{(\alpha_1)}_{\nu_{\alpha_1}},
\dots,\,N^{(\alpha_n)}_{\nu_{\alpha_n}}\}}=\int\limits_G
d\mu(g)\bar\chi^{(\Lambda)}(g)
[D^{(\alpha_1)}_{\nu_1\nu_1}]^{N^{(\alpha_1)}_{\nu_{\alpha_1}}}
\cdots
[D^{(\alpha_n)}_{\nu_n\nu_n}]^{N^{(\alpha_n)}_{\nu_{\alpha_n}}}
\end{equation}
is analytically extended for continuous values of variables
$N^{(\alpha_j)}_{\nu_{\alpha_j}}.$ This gives
\begin{multline}\label{derivP}
\frac{d\,\log \widetilde
\mathcal{P}^{\Lambda,\lambda_{\Lambda}}_{\{N^{(\alpha_1)}_{\nu_{\alpha_1}},
\dots,\,N^{(\alpha_n)}_{\nu_{\alpha_n}}\}}}
{d\,N^{(\alpha_j)}_{\nu_{\alpha_j}}}\\ = \frac{\int\limits_G
d\mu(g)\bar\chi^{(\Lambda)}(g)
[D^{(\alpha_1)}_{\nu_1\nu_1}]^{N^{(\alpha_1)}_{\nu_{\alpha_1}}} \cdots
[D^{(\alpha_n)}_{\nu_n\nu_n}]^{N^{(\alpha_n)}_{\nu_{\alpha_n}}}
\log[D^{(\alpha_j)}_{\nu_j\nu_j}]}{\int\limits_G d\mu(g)\bar\chi^{(\Lambda)}(g)
[D^{(\alpha_1)}_{\nu_1\nu_1}]^{N^{(\alpha_1)}_{\nu_{\alpha_1}}} \cdots
[D^{(\alpha_n)}_{\nu_n\nu_n}]^{N^{(\alpha_n)}_{\nu_{\alpha_n}}}}\,.
\end{multline}
Eq. (\ref{derivP}) give a set of conditions related to the internal symmetry
of a system. They are meaningful only for nonzero values of coefficients
(\ref{weights}). It is easy to see that coefficients
$\overline{P^{\Lambda,\lambda_{\Lambda}}_{\{N^{(\alpha_1)}_{\nu_{\alpha_1}},
\dots,\,N^{(\alpha_n)}_{\nu_{\alpha_n}}\}}}$ are different from zero only if
parameters $\lambda_{\Lambda}$ are consistent with the conservation of the
simultaneously measurable charges related to the symmetry group. A number of
such charges is equal to the rank $k$ of the symmetry group. For the isospin
$SU(2)$ group that is the third component of the isospin, for the flavor
$SU(3)$ that would be the third component of the isospin and the hypercharge.
In general case one has $k$ linear relations between variables
$N^{(\alpha_j)}_{\nu_{\alpha_j}}$ what reduces correspondingly the number of
independent variables.

\section{Isotopic Hadronic Gas}
Let consider a case of isotopic symmetry in a more detailed way. Diagonal
matrix elements for the representation $(j)$ in the Euler's angles
representation have the form \cite{Wigner, Edmonts}
\begin{subequations}
\begin{equation}
D^{(j)}_{mm}(\alpha,\beta,\gamma) =  \e^{im(\alpha+\gamma)}d^{(j)}_{mm}(\beta)
\ ,
\end{equation}
where
\begin{equation}
 d^{(j)}_{mm}(\cos\beta) =
\left(\frac{1+\cos\beta}{2}\right)^m P^{(0,\,2m)}_{j-m}(\cos\beta)\,,
\end{equation}
\end{subequations}
and $ P^{(0,\,2m)}_{j-m}(\cos\beta)$ are Jacobi polynomials
\begin{equation}\label{jacobi}
P^{(0,\,2m)}_{j-m}(x) = \frac{(-1)^{j-m}}{2^m
(j-m)!}(1+x)^{-2m}\frac{d^{j-m}}{dx^{j-m}}\left[(1-x)^{j-m}(1+m)^{j+m}\right]
\end{equation}
The measure $d\mu(g)$ for the $SU(2)$ group in this parametrization has the
form
\begin{eqnarray}
\int d\mu(g)f[g] ={1\over{8\pi^2}} \int\limits^{2\pi}_0d\alpha
\int\limits^{2\pi}_0d\gamma \int\limits^{\pi}_0d\beta\sin\beta
f[g(\alpha,\beta,\gamma)]\,.
\end{eqnarray}
There are only three possible nontrivial hadronic isotopic multiplets. These
are spinor $({\bf\frac{1}{2}})$, vector $(\bf 1)$, and $({\bf \frac{3}{2}})$
representations. Corresponding $d$ functions are

\begin{subequations}\label{dfunctions}
\begin{eqnarray}
   d^{(1/2)}_{{\pm 1/2},{\pm 1/2}}(\beta)& =&
\left(\frac{1+\cos\beta}{2}\right)^{\frac{1}{2}}\,,\\
   d^{(1)}_{\pm 1,\pm 1}(\beta)
&=&\frac{1+\cos\beta}{2}\,,\\
d^{(1)}_{0,0}(\beta) & = & \cos\beta\,,\\
d^{(3/2)}_{{\pm 3/2},{\pm 3/2}}(\beta)& = &
\left(\frac{1+\cos\beta}{2}\right)^{\frac{3}{2}}\,,\\
d^{(3/2)}_{{\pm 1/2},{\pm 1/2}}(\beta)& = & {\frac{1}{2}}{\left(\frac{1 +
\cos\beta}{2} \right)^{\frac{1}{2}}}( -1 + 3\cos\beta )
   \,.
  \end{eqnarray}
\end{subequations}

The group theoretic factor (\ref{otherP}) has the form

\begin{eqnarray}\label{weight1}
\lefteqn{\int\limits_G\,d\mu(g)\bar\chi^{(J)}(g)
[D^{(j_1)}_{m_{j_1}m_{j_1}}]^{N^{(j_1)}_{m_{j_1}}}\cdots
[D^{(j_n)}_{m_{j_n}m_{j_n}}]^{N^{(j_n)}_{m_{j_n}}}\nonumber}\\
&=&\frac{1}{8\pi^2}\sum_{M=-J}^J\int\limits_0^{2\pi}d\alpha\int\limits_0^{2\pi}d\gamma\int\limits_0^\pi
d\beta\sin\beta\,\e^{i(
N^{(j_1)}_{m_{j_1}}m_{j_1}+\cdots+N^{(j_n)}_{m_{j_n}}m_{j_n}-M)(\alpha+\gamma)}\nonumber\\
& &\times
d^{(J)}_{MM}(\cos\beta)\left[d^{(j_1)}_{m_{j_1}m_{j_1}}(\cos\beta)\right]^{N^{(j_1)}_{m_{j_1}}}
\cdots\left[d^{(j_n)}_{m_{j_n}m_{j_n}}(\cos\beta)\right]^{N^{(j_n)}_{m_{j_n}}}\,,
\end{eqnarray}
where $N^{(j)}_{m_j}$ is a number of particles with the isotopic coordinates
$\{j,m_j\}$.

The nonzero values are obtained only when the arguments of the exponent in Eq.
(\ref{weight1}) vanish
\begin{equation}\label{M}
N^{(j_1)}_{m_{j_1}}m_{j_1}+\cdots+N^{(j_n)}_{m_{j_n}}m_{j_n}-M=0
\end{equation}
For the hadronic system with the given value $\tilde M$ of the third component
of the isospin the factor (\ref{otherP}) is
\begin{eqnarray}\label{weight2}
\lefteqn{\int\limits_G\,d\mu(g)\bar\chi^{(J)}(g)
[D^{(j_1)}_{m_{j_1}m_{j_1}}]^{N^{(j_1)}_{m_{j_1}}}\cdots
[D^{(j_n)}_{m_{j_n}m_{j_n}}]^{N^{(j_n)}_{m_{j_n}}}\nonumber}\\
&=&\frac{1}{8\pi^2}\,\delta_{\tilde
M,\,N^{(j_1)}_{m_{j_1}}m_{j_1}+\cdots+N^{(j_n)}_{m_{j_n}}m_{j_n}}\int\limits_0^{2\pi}d\alpha\int\limits_0^{2\pi}d\gamma\int\limits_0^\pi
d\beta\sin\beta\,\nonumber\\
& &\times d^{(J)}_{\tilde M \tilde
M}(\cos\beta)\left[d^{(j_1)}_{m_{j_1}m_{j_1}}(\cos\beta)\right]^{N^{(j_1)}_{m_{j_1}}}
\cdots\left[d^{(j_n)}_{m_{j_n}m_{j_n}}(\cos\beta)\right]^{N^{(j_n)}_{m_{j_n}}}\,.
\end{eqnarray}

These equations allow to write general forms of the isotopic symmetry factor
(\ref{otherP}) for the hadronic system
\begin{equation}\label{genweight2}
\begin{split}
&\widetilde{\mathcal P}^{(J,\tilde
M)}_{\{N^{(1/2)}_{-1/2},N^{(1/2)}_{1/2},N^{(1)}_{-1},N^{(1)}_0,N^{(1)}_1,
N^{(3/2)}_{-3/2},N^{(3/2)}_{-1/2},N^{(3/2)}_{1/2},N^{(3/2)}_{3/2}\}}\\
 = &\frac{2J+1}{2}\int\limits_{-1}^1 dx \left(\frac{1+x}{2} \right)^{\tilde
M+{\mathcal R}}\left(\frac{-1+3x}{2} \right)^{N^{(3/2)}_{-1/2}+N^{(3/2)}_{1/2}}
x^{N^{(1)}_0} P^{(0,\,2\tilde M)}_{J-\tilde M}(x)\,,
\end{split}
\end{equation}
where
\begin{equation}\label{R}
{\mathcal R}=
 N^{(j_1)}_{m_{j_1}}|m_{j_1}|+\cdots+N^{(j_n)}_{m_{j_n}}|m_{j_n}|
\end{equation}
 The definition (\ref{R})
and the constraint (\ref{M}) allow to write
\begin{equation}\label{MplusR}
\mathcal{R}+\tilde M=2\sum_{\{j,|m_j|\}} N^{(j)}_{|m_j|}|m_j|
\end{equation}
 This is because
\begin{subequations}
\begin{equation}
\sum_{\{j,m_j\}} N^{(j)}_{m_j}m_j= \sum_{\{j,|m_j|\}}
N^{(j)}_{|m_j|}|m_j|-\sum_{\{j,|m_j|\}} N^{(j)}_{-|m_j|}|m_j|
\end{equation}
and
\begin{equation}
\sum_{\{j,m_j\}} N^{(j)}_{m_j}|m_j|= \sum_{\{j,|m_j|\}}
N^{(j)}_{|m_j|}|m_j|+\sum_{\{j,|m_j|\}} N^{(j)}_{-|m_j|}|m_j|
\end{equation}
\end{subequations}
This allows to write Eq. (\ref{genweight2}) in the form

\begin{multline}\label{genweight3}
\widetilde{\mathcal P}^{(J,\tilde
M)}_{\{N^{(1/2)}_{-1/2},N^{(1/2)}_{1/2},N^{(1)}_{-1},N^{(1)}_0,N^{(1)}_1,
N^{(3/2)}_{-3/2},N^{(3/2)}_{-1/2},N^{(3/2)}_{1/2},N^{(3/2)}_{3/2}\}}\\
 =  \frac{2J+1}{2}\int\limits_{-1}^1 dx \left(\frac{1+x}{2}
\right)^{(N^{(1/2)}_{1/2}+2N^{(1)}_1+N^{(3/2)}_{1/2}+3N^{(3/2)}_{3/2})}\\
\times\left(\frac{-1+3x}{2} \right)^{N^{(3/2)}_{-1/2}+N^{(3/2)}_{1/2}}
x^{N^{(1)}_0} P^{(0,\,2\tilde M)}_{J-\tilde M}(x)\,.
\end{multline}

One should notice that this reduction of independent variables is related only
to that part of the total weight (\ref{weights}) which comes from the group
theory. The condition (\ref{cond}) is obtained when the expression
(\ref{genweight3}) is multiplied by the combinatoric factor (\ref{faktor}).
Number of independent variables in the combinatoric factor can be reduced only
by conservation laws of corresponding charges of the system.

To be more specific let us consider as an example an iso-singlet system
consisting only of pions and nucleons \cite{TurRaf}. We take as independent
variables: $N$ - total number of particles, $B$ - total baryon number, $Q$ -
total electric charge, $ n_0$ - number of neutral pions, $n_N$ - number
of neutrons. Then \\

number of negative pions: $n_{-}=(N-n_N-n_0-Q)/2\,,$

number of positive pions: $n_{+}=-B+n_N+(N-n_N-n_0+Q)/2\,,$

number of protons: $n_P=B-n_N\,.$ \\

For iso-singlet state $B=2Q$ and
\begin{subequations}\label{partnumb}
\begin{eqnarray}
n_{-}& = &(N-n_N-n_0-Q)/2\,,\\
n_{+}& = &-2Q+n_N+(N-n_N-n_0+Q)/2\,,\\
n_P& = &2Q-n_N\,.
\end{eqnarray}
\end{subequations}
Let us assume \cite{TurRaf} that this system reaches the chemical equilibrium
in the evolution process governed by Vlasov - Boltzman kinetic equations with
the interactions restricted to binary collisions. Then the \textit{total}
number of particles remains constant but particles ratios are subjected to
constraints (\ref{cond}). For the given baryon number $B$ and the given total
number of particles $N,$ condition
\begin{equation}\label{condsingl}
\overline{P^{(0,\,0)}_{\{n_N,\,n_P,\,n_{-},\,n_0,\,n_{+}\} }}=\mbox{const.}
\end{equation}
gives evolution lines in the $n_N - n_0$ plane. The system evolves along these
lines which are here consequences of the isotopic $SU(2)$ symmetry and baryon
number conservation. The weights (\ref{condsingl}) expressed by means of
variables $B,\ N,\ n_N\ \mbox{and } n_0$ and calculated according to Eqs.
(\ref{faktorp}) and (\ref{partnumb}) have the form
\begin{eqnarray}\label{condexact}
\lefteqn{\overline{P^{(0,\,0)}_{\{n_N,\,n_P,\,n_{-},\,n_0,\,n_{+}\} }} =
\frac{B!}{2^B n_N!(B-n_N)!}}\\ &\times&\frac{(N-B)!}{3^{(N-B)} n_0!
(N/2-n_N/2-n_0/2-B/4)!(N/2+n_N/2-n_0/2-3 B/4)!}\nonumber \\
&\times&\frac{1}{2}\int\limits_{-1}^1 dx\,x^{n_0}
\left(\frac{1+x}{2}\right)^{(N-n_0-B/2)} \nonumber
\end{eqnarray}

\begin{figure}
\begin{center}{
{\epsfig{file=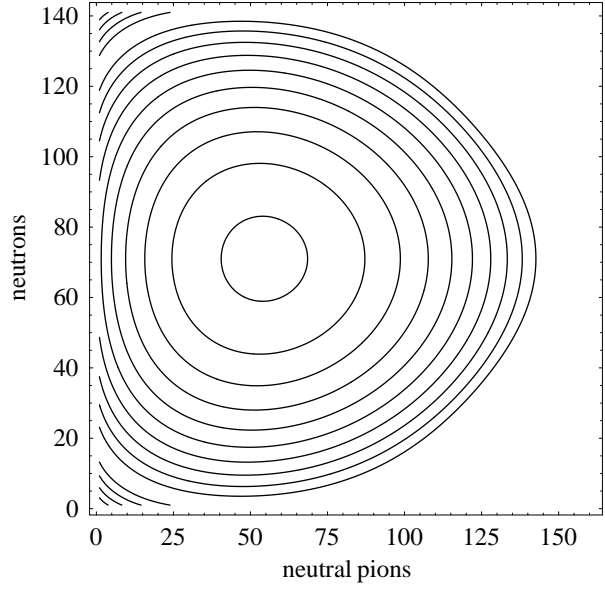,width=8cm}} }\end{center} \caption{Evolution curves for
the $\pi - N$ iso-singlet system consisting of 300 particles -- pions and
nucleons. Baryon number $B=140$.  The chemical equilibrium is reached along
equal weights curves (Eq. (\ref{condsingl})) in the plane $n_0 - n_N$. }
\label{Fig.1.}
\end{figure}

The corresponding evolution lines are shown in Fig. \ref{Fig.1.}.

\section{Conclusions}

We have got relations which are {\it necessary} global conditions
to provide consistency with the overall symmetry of the system.
They do not depend on the form of the underlying microscopic
interaction.  Abelian internal symmetries lead to simple and
obvious linear relations as in Eq.\,(\ref{abcharge}). Non-abelian
internal symmetries lead to nonlinear relations as in
Eq.\,(\ref{deriv}).

If we knew the solutions of symmetry invariant evolution equations then all
those constraints would become identities. In other case they give a subsidiary
information about the system and can be  used as a consistency check for
approximative calculations. A case of generalized Vlasov - Boltzman kinetic
equations was considered in \cite{TurRaf}.

New constraints lead to decreasing number of available states for the system
during its evolution. New correlations appear and the change in the
thermodynamical behavior can be expected. This can be observed in
hydrodynamical systems formed in high energy heavy ion collisions
\cite{ElRafTur}.

A challenging point is to find structures which would correspond to chemical
potentials when system approaches the equilibrium distribution. The equilibrium
distribution in the presence of constraints can be constructed by the Lagrange
multipliers method. The multipliers related to the ``abelian'' constraints,
such as Eq.\,(\ref{abcharge}), are well known chemical potentials. Multipliers
related to the ``non-abelian" constraints (\ref{deriv}) are more complicated.
Because these constraints are nonlinear ones, corresponding multipliers cannot
be treated as standard additive thermodynamical potentials.

\section*{Acknowledgments} Work supported in part by the Polish Committee for
Scientific Research under contract KBN~-~2P03B~030~18

\end{document}